\documentclass[a4paper,10pt,twoside]{cpc-hepnp}

\usepackage{multicol}
\usepackage{graphicx}
\usepackage{booktabs}
\usepackage{amssymb,bm,mathrsfs,bbm,amscd}
\usepackage[tbtags]{amsmath}
\usepackage{lastpage}
\usepackage{CJK}

\begin{document}
\begin{CJK*}{UTF8}{gbsn}

\fancyhead[c]{\small Chen Li-Zhu~ et al: High cumulants of conserved charges and their statistical uncertainties}


\title{High cumulants of conserved charges and their statistical uncertainties\thanks{This work was supported in part by the NSFC under Grant No. 11405088, No. 11521064, No. 11647093, the Major State Basic Research Development Program of China under Grant No. 2014CB845402,
and the Ministry of Science and Technology (MoST) under grant No. 2016YFE0104800.}}

\author{%
 Chen Li-Zhu(陈丽珠) $^{1;1)}$\email{chenlz@nuist.edu.cn}%
\quad Zhao Ye-Yin (赵烨印) $^{2}$
\quad Pan Xue(潘雪) $^{3}$
\quad Li Zhi-Ming(李志明) $^{2}$
\quad Wu Yuan-Fang(吴元芳)$^{2}$
}
\maketitle

\address{%
$^1$School of Physics and Optoelectronic Engineering, Nanjing University of Information Science and Technology, Nanjing 210044, China\\
$^2$Key Laboratory of Quark and Lepton Physics (MOE) and
Institute of Particle Physics, Central China Normal University, Wuhan 430079, China\\
$^3$School of Electronic Engineering, Chengdu Technological University, Chengdu 611730, China\\
}

\begin{abstract}
We study the influence of measured high cumulants of conserved charges on their associated statistical uncertainties in relativistic heavy-ion collisions. With a given number of events, the measured cumulants randomly fluctuate with an approximately normal distribution, while the estimated statistical uncertainties are found to be correlated with corresponding values of the obtained cumulants. Generally, with a given number of events, the larger the cumulants we measure, the larger the statistical uncertainties that are estimated.  The error-weighted averaged cumulants are dependent on statistics. Despite this effect, however, it is found that the three sigma rule of thumb is still applicable when the statistics are above one million.
\end{abstract}

\begin{keyword}
High cumulants, statistical uncertainty, statistics, QCD phase transition
\end{keyword}

\begin{pacs}
25.75.Gz, 25.75.Nq
\end{pacs}

\begin{multicols}{2}

\section{Introduction}

One of the main motivations of high energy heavy-ion collisions is to study the structure of the Quantum Chromodynamics (QCD)  phase diagram~\cite{STAR-QCD-1}.
The cumulants of conserved charges are powerful observables that are sensitive to the location of the QCD critical point~\cite{QCD-1, QCD-2, QCD-3, QCD-4, QCD-5, QCD-6, QCD-7}. Theoretical calculations predict that the third order cumulant ($C_3$) is proportional to correlation length ($\xi$) as $C_3\sim\xi^{4.5}$, and the fourth order cumulant grows more rapidly as $C_4\sim\xi^{7}$~\cite{QCD-1}.
Hence, the cumulants of conserved charges exhibit large fluctuations near the QCD critical point.
 Besides locating the QCD critical point, the extraction of the freeze-out temperature and baryon-chemical potential directly from QCD first principles becomes achievable,  by comparing the experimental and lattice QCD measured cumulants~\cite{freeze-out-qcd-1, freeze-out-qcd-2,freeze-out-qcd-3,freeze-out-qcd-4}.

With the data collected from the first beam energy scan (BES I) program at the Relativistic Heavy Ion Collider (RHIC), cumulants of net-proton~\cite{STAR-proton, STAR-proton-preliminary} and net-charge~\cite{STAR-charge, PHENIX-charge} multiplicity distributions have been measured.
For $\kappa\sigma^2$ of net-charge multiplicity distributions, where $\kappa\sigma^2\equiv C_4/C_2$, no energy dependence has been observed by either the STAR or PHENIX experiment.  The  $\kappa\sigma^2$ of net-proton multiplicity distributions, especially for the recent preliminary results with larger transverse momentum range, $0.4 < p_T < 2.0$ GeV/$c$, shows a possible signal of non-monotonic variation at RHIC/STAR.
 When comparing  experimental measurements to theoretical calculations, we should keep in mind that the non-critical background contribution can affect the measured results in the experiment. Currently, there are many discussions on techniques for
 reducing the non-critical contributions for cumulants, such as statistical error estimation~\cite{delta-theorem,xiaofeng-JPG-2013, bootstrap}, centrality bin width effect~\cite{cbwc-luo-1,cbwc-lizhu-1,cbwc-lizhu-C6}, and efficiency correction~\cite{koch-eff-1,koch-eff-2,luo-eff,Japan-eff-1}.

 With the current statistics collected at RHIC BES I, errors of the high cumulants are significantly larger than other event-by-event observables, such as dynamical net-charge fluctuations~\cite{dynamical-net-charge}, $K/\pi, p/\pi$ and $K/p$ fluctuations~\cite{STAR-kpi}. It has been found that the estimated errors are reasonable based on delta theorem and bootstrap methods~\cite{delta-theorem, bootstrap}.
 However, with limited statistics, we did not notice the relationship between the measured values and their corresponding errors.
 In fact, a study using the method of centrality bin width correction (CBWC) has indicated that the estimated error is influenced by the measured cumulant~\cite{xiaofeng-JPG-2013}.
 When applying the CBWC method with two different schemes, weighting the cumulant by the number of events or error,
 Ref. \cite{xiaofeng-JPG-2013} shows that values of $\kappa\sigma^2$ based on these two schemes are not consistent with each other. It indicates that the estimated error is not only proportional to $\sqrt{\frac{1}{n}}$, where $n$ is the number of events, but also to the measured value of $\kappa\sigma^2$.

\begin {figure*}[htp]
\begin{center}
\includegraphics[width=5.0in]{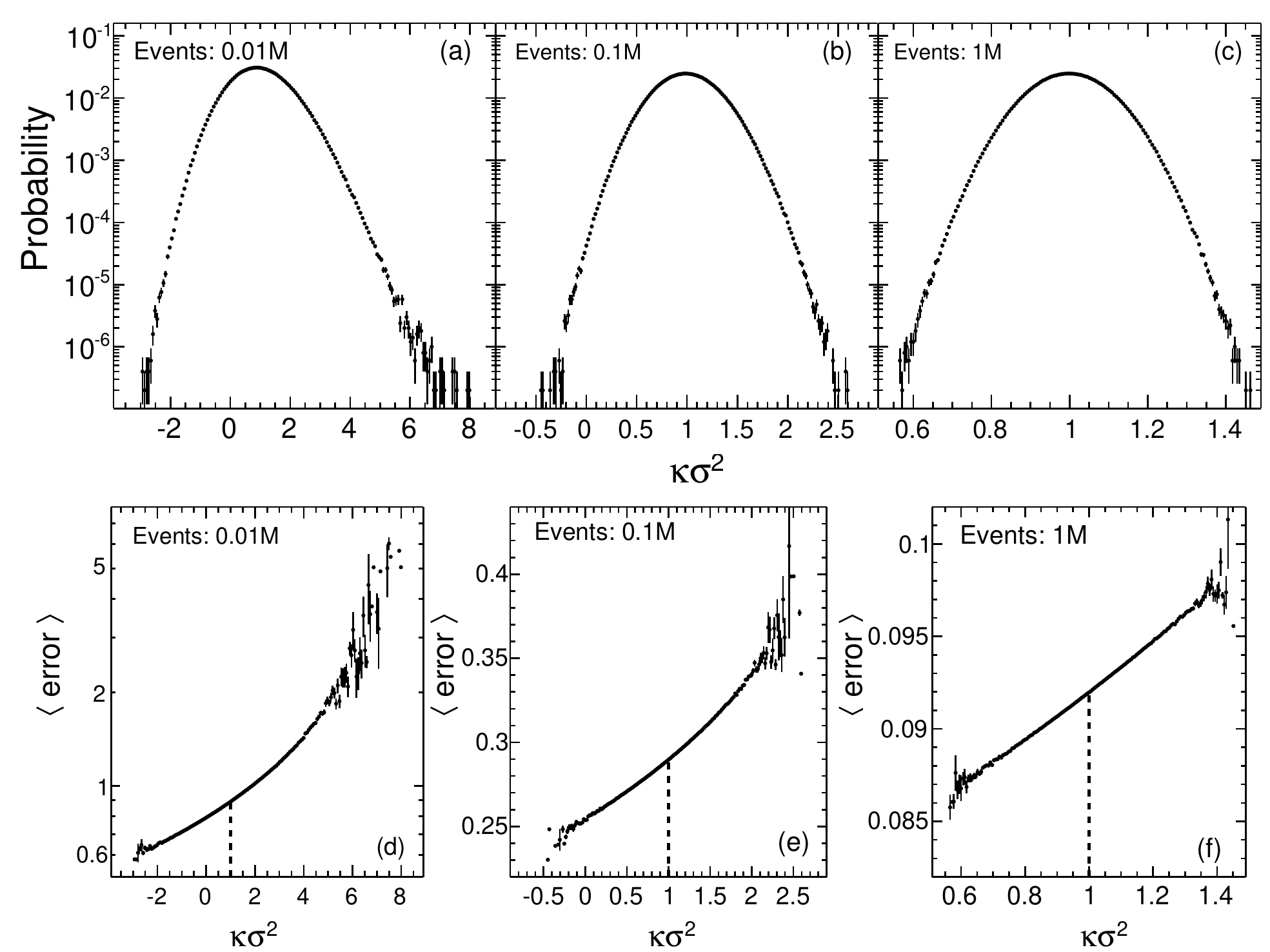}
\vspace{4pt}
\figcaption{\label{error-correlation} The upper panel shows probability distributions of $\kappa\sigma^2$ with different statistics. The lower panel shows the averaged error of $\kappa\sigma^2$, $\left<error\right>$, as a function of its measured $\kappa\sigma^2$. The vertical dashed lines show the values of $\left<error\right>$ when $\kappa\sigma^2$ is unity.}
\end{center}
\end{figure*}

For $N$ independent observations  ($x_1, x_2, \cdots, x_N$)  with the same unknown expectation ($\mu$), they are approximated with a normal distribution as $x_i\sim(\mu, \sigma^2)$, if they are obtained from independently generated samples with the same statistics and conditions. The error of $x_i$ is the width of this normal distribution ($\sigma$). Experimentally, we cannot directly obtain the width $\sigma$ from the normal distributions, since what we can obtain is just one measurement of the $N$ observations. In this case, the same data sample is utilized to obtain the value of the measurement and estimate the corresponding error.  Consequently, the estimated error is influenced by the measured value. In this paper, we will comprehensively study the properties of errors of $\kappa\sigma^2$ in relativistic heavy-ion collisions. We will start with discussions of the correlation between measured values of  $\kappa\sigma^2$ and estimated uncertainties. Generally, with a given number of events, the larger $\kappa\sigma^2$, the larger the statistical uncertainties that are estimated. In Section 3, we will discuss the effectiveness of the measured averaged $\kappa\sigma^2$  in two schemes: the event- and error- weighted.
The probabilities of $\kappa\sigma^2$ that lie outside three and four standard deviations of the expectation are discussed in Section 4.
Finally, the results are summarized in Section 5.

\section{Estimated uncertainty of $\kappa\sigma^2$}

Supposing the particle and anti-particle are produced independently according to a Poisson distribution, then the net-particle follows a Skellam distribution~\cite{skellam-1,skellam-2,skellam-3}. The parameters of the Skellam distribution are the means of those for the particles and anti-particles.  Referring to the means of protons and anti-protons at low energies of RHIC/STAR, the parameters of the Skellam distribution are set to $m_1=14.5, m_2=0.6$~\cite{STAR-proton}.

\begin {figure*}[htbp]
\begin{center}
\includegraphics[width=4.0in]{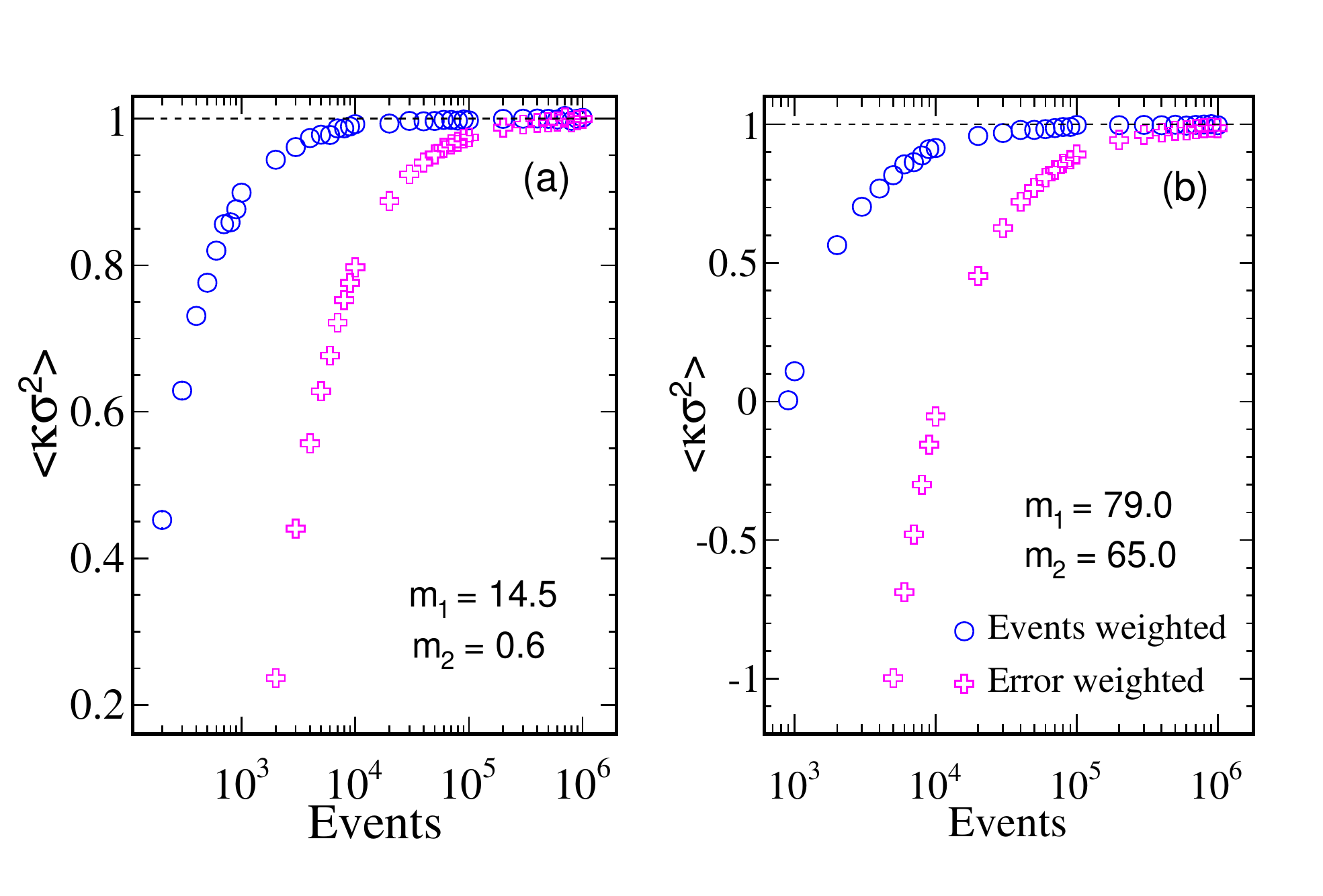}
\caption{(Color online) Statistics dependence of $\left<\kappa\sigma^2\right>$ obtained by event- and error- weighted averages, respectively. The results are extracted from  simulations of the Skellam
distribution with input parameters (a) $m_1$ = 14.5
and $m_2$ = 0.6, and (b) $m_1$ = 79.0 and $m_2$ = 65.0. These parameters can be referenced to net-proton and net-charge multiplicity distributions for moment analysis at RHIC/STAR.
\label{expectation-weight}}
\end{center}
\end{figure*}

The upper panel of Fig.~\ref{error-correlation} shows probability distributions of $\kappa\sigma^2$ with different numbers of events: 0.01 million (0.01M) , 0.1M and 1M. In each sub-figure, the probability distribution is obtained from 5M independent randomly generated samples. The means of these three distributions are 0.9895, 0.9989 and 0.9998, respectively. Those values are on the brink of unity, which is the theoretical expectation.
 The widths of these three distributions are 0.91,  0.29 and 0.092, respectively.
 According to the definition of the statistical error, the width of the distribution is the statistical error with the given statistics.
 In our simulations,  the same data are used to measure $\kappa\sigma^2$ and estimate their errors. Consequently, in Fig.~\ref{error-correlation}(a), the estimated errors of all $\kappa\sigma^2$ are of course not completely the same as the widths of the distributions. Not only that, the estimated error should be different and fluctuate even for the same values of $\kappa\sigma^2$.
 To generally see the correlation between the values of $\kappa\sigma^2$ and their associated errors,
Figs.~\ref{error-correlation}(d) to (f) show $\left<error\right>$ as a function of $\kappa\sigma^2$.
$\left<error\right>$ is the average of the estimated errors of $\kappa\sigma^2$ in the same histogram bin.
The binning method of $\kappa\sigma^2$  in $x-$axis in the lower panel is exactly the same as that in the upper panel.
With the given statistics, the larger the cumulant we measured, the larger the $\left<error\right>$ we obtained. $\left<error\right>$ is influenced by the measured $\kappa\sigma^2$.

When $\kappa\sigma^2$ is unity, the vertical lines from Figs.~\ref{error-correlation}(d) to (f) show the values of $\left<error\right>$ are 0.91,  0.29 and 0.092, respectively. Those values are in agreement with the widths of the corresponding probability distributions in the upper panel. Consequently, $\left<error\right>$ is under-estimated when $\kappa\sigma^2$ is smaller than its expectation, while $\left<error\right>$ is over-estimated when $\kappa\sigma^2$ is larger than its expectation.
As the statistics increase, Figs.~\ref{error-correlation}(d) to (f)  demonstrate that the increment of $\left<error\right>$ decreases.
 Consequently, the larger the statistics we use to measure $\kappa\sigma^2$,  the smaller the correlation between the measured $\kappa\sigma^2$ and its estimated statistical error.

Here we only show influence of the measured $\kappa\sigma^2$ on its statistical error.
We want to give a reminder that this kind of relationship cannot be eliminated and it is not dependent on the details of the method of error propagation as long as the same collected data  are used to extract the measured value and its corresponding error.

Since this influence cannot be ignored in cumulant analysis, we have to be careful when extracting some other results based on cumulants and their errors. Experimentally,  the rule of  three sigma (or even higher) standard deviations is the most important indication of a new phenomenon, where $\sigma$ is the error of measurement. Currently, in order to reduce the initial size fluctuation, it is suggested to calculate the cumulant at each of $N_{ch}$. The cumulant is averaged over all multiplicities in a given centrality, where the average is always weighted by the number of events in each $N_{ch}$. This is called the centrality bin width correction ~\cite{xiaofeng-JPG-2013, xiaofeng-star-cbwc}. The difference in the cumulant obtained by the CBWC method based on error-weighted average and event-number average should be studied.
In the following sections, we will discuss  $\kappa\sigma^2$ obtained by these two different schemes.

\section{Two schemes to evaluate $\left<\kappa\sigma^2\right>$}

For $N$ independent observations ($x_1, x_2, \cdots, x_N$) with the same unknown expectation ($\mu$),  if the error of each observable is $e_i$, the maximum likelihood estimations of their expectation and error~\cite{maximum} are

\begin {equation}\label{mean1}
\left<x\right> = \frac{\sum_{i=1}^{i=N} x_i/e_i^2}{\sum_{i=1}^{i=N} 1/e_i^2},
\end {equation}
and
\begin {equation}\label{error1}
error = \sqrt{\frac{1}{\sum_{i=1}^{i=N} 1/e_i^2}}
\end {equation}

\begin {figure*}[htbp]
\begin{center}
\vspace{4pt}
\includegraphics[width=4.6in]{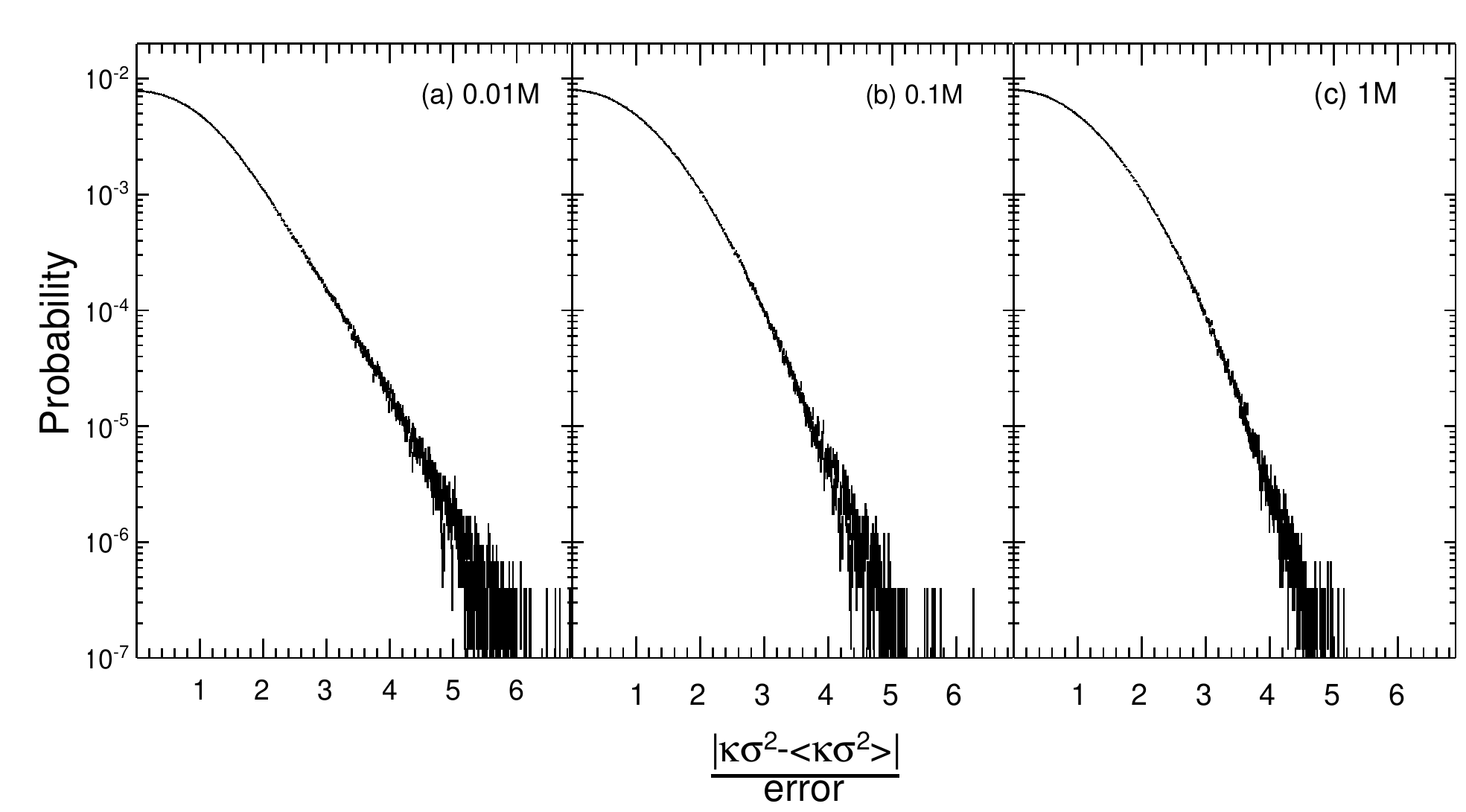}
\figcaption{ Probability distributions of $\frac{|\kappa\sigma^2-\left<\kappa\sigma^2\right>|}{error}$  with different statistics.
\label{sigma-deviation}}
\end{center}
\end{figure*}

In this case, the expectation is evaluated by the error-weighted average.  In general, we can choose this method to obtain $\left<x\right>$. However, the same data are always used to obtain $x_i$ and its error $e_i$ in experiment, which will lead to a correlation between $x_i$ and $e_i$.  In theory, if  the difference of $e_i$ is only proportional to  $\sqrt{\frac{1}{n_i}}$,
their expectation and error can be determined by weighting the average by the number of events,

\begin {equation}\label{mean2}
\left< x\right> = \frac{\sum_{i=1}^{i=N} x_in_i}{\sum_{i=1}^{i=N} n_i},
\end {equation}
and
\begin {equation}\label{error2}
error = \frac{\sqrt{\sum_{i=1}^{i=N} \left(e_in_i\right)^2}}{\sum_{i=1}^{i=N} n_i}.
\end {equation}

 Fig.~\ref{expectation-weight} demonstrates the statistics dependence of  $\left<\kappa \sigma^2\right>$ obtained by event- and error- weighted averages, respectively. $\left<\kappa \sigma^2\right>$ first increases as the number of events increases based on these two schemes. As we have mentioned in Ref.~\cite{chenlz-distribution}, $\kappa\sigma^2$ is sensitive to the tail of the net-particle multiplicity distributions.  When the number of events is insufficient, the statistics are lower, and the information we can detect about the tail of the net-particle multiplicity distributions is poorer.  That is why $\left<\kappa \sigma^2\right>$ obtained from the event-weighted average firstly increases as the number of events increases, and then finally saturates to unity.

Fig.~\ref{expectation-weight}  shows that the effect of underestimation for $\left<\kappa \sigma^2\right>$ obtained from the error-weighted average is stronger than that obtained from the event-weighted average. The minimum statistics required for the event-weighted method is around 0.01M events, while it is about 0.1M events for the error-weighted method. Currently, in most central collisions (0-5\% centrality), the number of events in each of $N_{ch}$, or in each $\delta1\%$ centrality bin width, is significantly lower than 0.1M. This is why the event-weighted average is in favor in current RHIC/STAR experiments~\cite{STAR-proton, STAR-charge,cbwc-luo-1}. On the other hand, we still need to know that the event-weighted method is effective only when the number of events in each of $N_{ch}$ is larger than 0.01M. Currently, the analyzed statistics for cumulants of net-proton multiplicity distributions is 6.6M at $\sqrt{s_{NN}}$ = 11.5 GeV in Au + Au collisions. If we simply assume that $\frac{1}{16}$ of events are within 0-5\% centrality and there are 100 multiplicity bins in 0-5\% centrality, then the number of events in each $N_{ch}$ is just around 4100. It means that $\kappa\sigma^2$ would be under-estimated if the event-weighted CBWC method is applied in each of $N_{ch}$ at $\sqrt{s_{NN}}\le$ 11.5 GeV.

Fig.~\ref{expectation-weight}(b) shows qualitatively similar statistical dependence to Fig.~\ref{expectation-weight}(a). However, it requires more statistics to approach the saturated value of unity. This means more statistics are needed to get a stable expectation as the parameters of the Skellam distribution become larger.
So the required statistics depend on the distribution, or the mechanism of particle production.

Here, we only show the statistics dependence of $\left<\kappa \sigma^2\right>$. For other cumulants, such as $S\sigma$ and $C_6/C_2$,
the required statistics should be observable dependent.
It should be studied carefully case by case.

\section{Deviation of $\kappa\sigma^2$ from its expectation}
Since the estimated error is influenced by the measured $\kappa\sigma^2$, we should examine that if three-sigma rule of thumb is still applicable.
To see the deviation of measured $\kappa\sigma^2$ from its expectation, the probability distributions of $\frac{|\kappa\sigma^2-\left<\kappa\sigma^2\right>|}{error}$, $P\left(\frac{|\kappa\sigma^2-\left<\kappa\sigma^2\right>|}{error}\right)$ , is  demonstrated in Fig.~\ref{sigma-deviation}.
With three cases of statistics shown in Fig.~\ref{sigma-deviation},
$P\left(\frac{|\kappa\sigma^2-\left<\kappa\sigma^2\right>|}{error}\right)$ decreases rapidly as $\frac{|\kappa\sigma^2-\left<\kappa\sigma^2\right>|}{error}$ increases.
The tail of $P\left(\frac{|\kappa\sigma^2-\left<\kappa\sigma^2\right>|}{error}\right)$ is dependent on the statistics. The lower the statistics, the longer the tail of the probability distribution.

In a normal distribution, 0.3\% and 0.007\% of the values lie outside $3\sigma$ and $4\sigma$  standard deviations of the expectation respectively.
Table~\ref{3sigma-deviation} shows that when statistics are lower than 1M, the values of $P\left(\frac{|\kappa\sigma^2-\left<\kappa\sigma^2\right>|}{error}>3.0\right)$ and $P\left(\frac{|\kappa\sigma^2-\left<\kappa\sigma^2\right>|}{error}>4.0\right)$ are all slightly larger than that in a normal distribution.
Therefore,  three sigma and/or four sigma confidence intervals for $\kappa\sigma^2$ are applicable when the number of the events is above 1M.

\begin{center}
\tabcaption{\label{3sigma-deviation} Probabilities of measured $\kappa\sigma^2$ that lie outside three and four standard deviations of the expectation.}
\footnotesize
\begin{tabular*}{85mm}{c@{\extracolsep{\fill}}cccccc}
\hline
Statistics (in millions)  &  0.01 & 0.1 & 1\\
\hline
$P\left(\frac{|\kappa\sigma^2-\left<\kappa\sigma^2\right>|}{error}>3.0\right)$ (\%)  & 0.74&0.34& 0.28 \\
\hline
$P\left(\frac{|\kappa\sigma^2-\left<\kappa\sigma^2\right>|}{error}>4.0\right)$ (\%)  & 0.079&0.015& 0.0077\\
\bottomrule
\end{tabular*}
\end{center}

\section{Summary}

In summary, we have studied high cumulants of conserved charges and their statistical uncertainties in relativistic heavy-ion collisions. With a given number of events in Monto Carlo simulations, the measured cumulants randomly fluctuate with an approximately normal distribution. The mean of the distribution is equal to the theoretical expectation and the width of the distribution is proportional to $\sqrt{\frac{1}{n}}$. The estimated uncertainty is not only determined by $\sqrt{\frac{1}{n}}$, but also influenced by the measured values of the cumulants.
Generally, with a given number of events, the larger the cumulants measured, the larger the estimated statistical uncertainties.  As the number of events increases, the influence of the measured cumulant on its statistical uncertainty is reduced.

Since the estimated statistical uncertainties are influenced by their measured values of cumulant, we should be careful if we re-evaluate other measurements based on the measured cumulants and their uncertainties. It is found that values of $\left<\kappa\sigma^2\right>$ are different depending on whether the event-weighted or error-weighted method is used. Consequently, the values of $\kappa\sigma^2$ are different when using the CBWC method based on the event-weighted and error-weighted methods. With limited available STAR BES I data, the error-weighted average scheme is not suitable. At $\sqrt{s_{NN}}\le$ 11.5 GeV, in most central collisions, $\kappa\sigma^2$ is also under-estimated if the event-weighted CBWC method is applied in each of $N_{ch}$.

We have also studied the probability distributions of $\frac{|\kappa\sigma^2-\left<\kappa\sigma^2\right>|}{error}$ with different statistics, and found that the three sigma rule of thumb is applicable when the statistics are above 1M.

\end{multicols}

\vspace{-1mm}
\centerline{\rule{80mm}{0.1pt}}
\vspace{2mm}

\begin{multicols}{2}

\end{multicols}

\clearpage

\end{CJK*}
\end{document}